\documentclass[]{iopart}
\usepackage{graphicx}
\usepackage{subfigure}
\usepackage{dcolumn}
\usepackage{bbm}
\usepackage{color}
\usepackage{amscd}

\begin{document}
\title{Signatures of non-Markovianity in classical single-time 
probability distributions}

\author{A Smirne, A Stabile and B Vacchini}

\address{Dipartimento di Fisica, Universit{\`a} degli Studi di
Milano, Via Celoria 16, I-20133 Milan, Italy\\
INFN, Sezione di Milano, Via Celoria 16, I-20133
Milan, Italy}

\ead{andrea.smirne@unimi.it}

\begin{abstract}
We show that the Kolmogorov distance allows to quantify memory effects in classical stochastic processes
by studying the evolution of the single-time probability distribution. We further investigate the relation
between the Kolmogorov distance and other sufficient but not necessary
signatures of non-Markovianity within the classical setting.

\end{abstract}

\section{Introduction}

In the theory of classical stochastic processes the definition
of Markov process relies on a requirement involving the entire
hierarchy of conditional probability distributions \cite{Feller1971}. A 
non-Markov process is then defined to be a stochastic process whose 
conditional
probabilities do not comply with this requirement. 
In particular, the transition probabilities
of a Markov process obey the well-known Chapman-Kolmogorov
equation, which in the homogeneous case
can be read as a semigroup composition law for a family of transition 
operators
defined in terms of the transition probabilities of the process 
\cite{Feller1971}.
A Markov process can be considered as a process that lacks memory,
whereas non-Markov processes describe phenomena
in which memory effects are relevant.

Different approaches to the notion
of stochastic process in quantum mechanics have been
developed, based on $C^*$-algebras \cite{Lindblad1979, Lewis1981, 
Accardi1982} or on the stochastic Schr{\"o}dinger equation 
\cite{Diosi1988, Barchielli1991, Gardiner2004},
just to mention some relevant examples,
but there is not a  well-established general theory of quantum 
stochastic processes.
The crucial difference between the quantum and the classical setting 
traces
back to the peculiar role played by measurement in the quantum
case. In order to make statements
about the values of observables of a quantum system at different times 
one has to
specify measurement procedures, which modify the subsequent evolution 
\cite{Holevo2001}.
The generic dynamics of a quantum system,
also taking the interaction with its environment into account, i.e.
the dynamics of an open quantum system 
\cite{Breuer2002}, is usually formulated by means of a one-parameter
family of completely positive dynamical maps,
which fixes the evolution of the statistical
operator representing the open system's state. Within this framework, 
where a straightforward translation of the classical definition does not apply,
the non-Markovianity of a quantum dynamics has recently been defined and 
quantified 
in terms of specific properties of the dynamical maps \cite{Wolf2008, Breuer2009, Rivas2010, Breuer2012}.
In analogy  with the transition operators of classical Markov processes,
one can identify quantum Markovianity with a proper composition law of 
the dynamical maps \cite{Rivas2010}, which reduces to
the semigroup law in the time-homogeneous case. 
In a different and non-equivalent way \cite{Laine2010, Haikka2011}, one 
can identify quantum Markovianity
with the contractivity of the trace distance 
between states of the open quantum system during the dynamics \cite{Breuer2009}. 
The trace distance quantifies the distinguishability of quantum states 
\cite{Nielsen2000}, so that its variations in the course of time can be 
read
in terms of an information flow between the open system and 
its environment \cite{Breuer2009, Laine2010, Breuer2012}.
An increase of the trace distance indicates that the information is 
flowing from the
environment to the open system, so that the influence the open system 
has had on
the environment affects the open system back again. 
The trace distance allows 
to formulate the physical idea of memory effects as typical of 
non-Markovian dynamics in a mathematically definite
way.

The precise relation between the definition 
of Markov stochastic process and the different notions of quantum 
Markovianity has been investigated in \cite{Vacchini2011, Vacchini2012}. 
More specifically, one can introduce also at the classical level
a family of dynamical maps which describes the evolution of
the single-time probability distribution of a stochastic process.
The properties used to define the Markovianity of quantum dynamics 
naturally induce 
signatures of non-Markovianity for the classical stochastic processes.
The evolution of the single-time probability distribution does not allow 
to assess
the Markovianity of a stochastic process.
However, the violation of either of the above-mentioned properties 
provides a
sufficient condition for the stochastic
process to be non-Markovian \cite{Vacchini2011, Vacchini2012}. In this 
paper, we deepen the analysis
of the signatures of non-Markovianity and our entire 
discussion will be kept within the classical setting. 
We first focus on the signature of non-Markovianity
based on the evolution of the distinguishability between
single-time probability distributions, which is quantified through the 
Kolmogorov distance, i.e., the classical
counterpart of the trace distance. 
In particular, we show that the increase in time of the Kolmogorov 
distance provides a quantitative description of
the memory effects present in the evolution of single-time probability
distributions. We take into account two-site semi-Markov processes 
\cite{Feller1971},
which allow for a complete characterization in terms of a waiting time 
distribution.
By means of the Kolmogorov distance, we study how the different features
of the waiting time distribution influence the memory effects in the 
subsequent evolution 
and then the possibility to detect
the non-Markovianity of the process at the level of single-time 
probability distributions.
Furthermore, we investigate the relation between the Kolmogorov distance 
and the
differential equation satisfied by the single-time probability 
distribution.
The latter is strictly related to the composition law of the classical 
dynamical maps 
and it provides a further signature of non-Markovianity,
which is shown not to be equivalent to the increase of the Kolmogorov distance.

\section{Memory effects in semi-Markov processes}

For the sake of simplicity, let us consider a stochastic process
taking values in a finite set. The single-time probability distribution 
at time $t$ is then a
probability vector ${\bi{p}} \left( t \right)$, so that its elements 
denoted 
as $p_k(t)$ $k=1, \ldots N$, satisfy $p_k (t) \geq 0$ and $\sum_k p_k(t) 
=1$. 
As in the quantum setting \cite{Breuer2002}, one can
describe the time evolution of ${\bi{p}} \left( t \right)$ by means of a 
one-parameter family
of dynamical maps $\left\{\Lambda(t, 0)\right\}_{t \geq 0}$
\begin{eqnarray}
  {\bi{p}} \left( t \right) & = & \Lambda \left( t, 0\right) {\bi{p}}
  \left( 0 \right),   \label{eq:cmap}
\end{eqnarray}
with $t_0=0$ fixed initial time.
It is easy to see that a matrix $\Lambda$ associates probability vectors 
to probability vectors
if and only if its entries $(\Lambda)_{j k}$ satisfy the conditions:
\begin{equation}
\left( \Lambda 
\right)_{j k} \geq 0 \qquad
\sum^N_{j=1} \left( \Lambda 
\right)_{j k} =1, \label{eq:stochmatcond}
\end{equation}
$\forall j, k = 1, \ldots, N $.
A matrix which fulfills equation~(\ref{eq:stochmatcond}) is usually 
called stochastic matrix.
We will thus require that every dynamical map $\Lambda(t, 0)$ is a 
stochastic matrix.

As long as only the single-time probability distribution is taken into 
account, one cannot in general
determine whether the stochastic process is Markovian or not. 
Nevertheless,
one can introduce signatures of non-Markovianity at the level of the 
evolution of single-time probability
distributions. These can be understood as sufficient but not necessary 
conditions
for the process to be non-Markovian. The Kolmogorov
distance $D_K \left({\bi{p}}^1, {\bi{p}}^2\right)$ between  two 
probability distributions ${\bi{p}}^1$ and ${\bi{p}}^2$
on a common finite set $\mathcal{X}$ is defined as
\begin{eqnarray}
  D_K \left({\bi{p}}^1, {\bi{p}}^2\right) & = & \frac{1}{2} \sum_{k\in \mathcal{X}} 
\left| p^1_ k - p^2_k
\right|.  \label{eq:kdist}
\end{eqnarray}
Now, consider the time evolution of the Kolmogorov distance $D_K 
\left({\bi{p}}^1(t), {\bi{p}}^2(t)\right)$
between probability vectors evolved from different initial conditions 
through the stochastic matrix in (\ref{eq:cmap}).
The Chapman-Kolmogorov equation implies \cite{Vacchini2011} $D_K 
\left({\bi{p}}^1(t), {\bi{p}}^2(t)\right) < D_K \left({\bi{p}}^1(s), 
{\bi{p}}^2(s)\right)$, $\forall t\geq s$,
so that any revival in the evolution of the Kolmogorov distance is a 
signature of the non-Markovianity of the process.
By adapting the interpretation introduced for the dynamics of open 
quantum
systems \cite{Breuer2009, Breuer2012}, any increase of the Kolmogorov 
distance
can be read as a memory effect in the evolution of the single-time 
probability distribution.
The information about the initial condition that is contained in the 
initial value
of the Kolmogorov distance can be partially lost and then recovered, 
implying
an increase of the distinguishability between the probability 
distributions evolved
from the two different initial conditions. We can quantify
the memory effects in the overall dynamics fixed by the stochastic matrices $\Lambda(t,0)$
by integrating the rate of change of the Kolmogorov distance 
$\sigma(t,{\bi{p}}^{1,2}(0)) =(\rmd/ \rmd t) D_K 
\left({\bi{p}}^1(t), {\bi{p}}^2(t)\right)$
over the time-intervals where the information about the initial 
condition is recovered.
Namely, we define (compare with \cite{Breuer2009})
\begin{equation}\label{eq:miscl}
\mathcal{N}_C(\Lambda) = \max_{{\bi{p}}^{1,2}(0)} \int_{\sigma > 0} \rmd t \, 
\sigma(t,{\bi{p}}^{1,2}(0)),
\end{equation}
where the maximum among the couples of initial probability vectors is 
taken.
$\mathcal{N}_C(\Lambda)$ points out the relevance of the signature 
of non-Markovianity in the evolution of 
single-time probability distributions
given by the increase of the Kolmogorov distance.

As a specific example, let us consider the semi-Markov processes 
\cite{Feller1971},
which are a class of stochastic processes allowing for a compact 
characterization.
The semi-Markov processes combine features of Markov chains and renewal 
processes \cite{Ross2007}:
they describe a system moving among different sites in a way such that 
the random times separating the
different transitions as well as the transition probabilities between 
the different sites only depend
on departure and arrival sites.  A semi-Markov process is completely 
determined by
the semi-Markov matrix $Q(t)$, whose entries $(Q)_{j k}(t)$ are the 
probability densities to make the jump $k\rightarrow j$
in a time $t$. For the sake of simplicity, we will take into account a 
system
moving between two different sites and a semi-Markov matrix $Q(t) 
=\left(\begin{array}{cc}
    0 & 1 \\
    1& 0 
  \end{array}\right) f(t)$. This means that once in a site the system
jumps with certainty to the other, with a random time before
each new jump given by the site-independent waiting time distribution 
$f(t)$.
The semi-Markov process will be Markovian according to the precise 
definition
of the classical stochastic processes if and only if the waiting time
distribution is the exponential one \cite{Breuer2008}
\begin{equation}\label{eq:exp}
g(t) \equiv \lambda e^{-\lambda t}.
\end{equation}
For a generic waiting time distribution,
the dynamical maps $\Lambda(t,0)$, which can be obtained
through an integrodifferential equation fixed by the semi-Markov matrix 
\cite{Vacchini2011}, are
\begin{equation}
\Lambda(t, 0)= \frac{1}{2}\left(\begin{array}{cc}
    1+ q(t) & 1- q(t) \\
    1- q(t)& 1+q(t) 
  \end{array}\right).\label{eq:kdp}
\end{equation}
Here, $q \left( t \right) = \sum^{\infty}_{n = 0} p (2n, t) -
  \sum^{\infty}_{n = 0} p(2n+1, t)$ expresses the difference between the 
probability to have an even or
an odd number of jumps up to time $t$. This quantity is related  to the 
waiting time distribution
via the corresponding Laplace transforms
\begin{eqnarray}
  \hat{q} \left( u \right) & = & \frac{1}{u}  \frac{1 - \hat{f} \left( u
  \right)}{1 + \hat{f} \left( u \right)},  \label{eq:qu}
\end{eqnarray}
where $ \hat{v} (u) =  \int^{\infty}_0 \rmd t\, v(t)\, e^{- u t}$ is 
the Laplace transform of the function $v(t)$.
The dynamical map $\Lambda(t, 0)$ associates the initial probability 
vector to the probability vector at time $t$ and then determines
the expression of the Kolmogorov distance. In particular, for 
$\Lambda(t, 0)$ as in (\ref{eq:kdp})
one has 
\begin{equation}\label{kkdd}
D_K \left({\bi{p}}^1(t), {\bi{p}}^2(t)\right) = |q(t)| D_K 
\left({\bi{p}}^1(0), {\bi{p}}^2(0)\right).
\end{equation}
Thus, the memory effects in the evolution of the single-time probability 
distribution
are quantified by, see (\ref{eq:miscl}), 
\begin{equation}\label{eqq}
\mathcal{N}_C(\Lambda) =\int_{\Omega_+} \rmd t \frac{\rmd}{\rmd t} |q(t)| = 
\sum_i\left(|q(b_i)|-|q(a_i)|\right),
\end{equation}
where $\Omega_+= \bigcup_i (a_i, b_i)$ corresponds to the time intervals 
in which $|q(t)|$ increases.
The maximum growth of the Kolmogorov distance in the course of time is 
obtained for the two initial probability
vectors $(1,0)^{\top}$ and $(0,1)^{\top}$.

By studying the quantity $\mathcal{N}_C(\Lambda)$ in (\ref{eqq}) for different 
choices
of the waiting time distribution $f(t)$, we can describe how different 
semi-Markov processes are characterized
by a different amount of memory effects in the evolution of 
${\bi{p}}(t)$.
An interesting class of waiting time distributions is given by the 
so-called special Erlang distributions of order $n$.
The latter consist in the convolution of $n$ equal exponential waiting 
time distributions of parameter $\lambda$, see (\ref{eq:exp}):
\begin{equation} \label{conv}
f(t) = \underbrace{(g\ast \dots g)}_{n}(t).
\end{equation}
A special Erlang distribution of order $n$ describes a random variable 
given
by the sum of $n$ independent and identically distributed exponential 
random variables. This waiting time distribution can be related to
a system that jumps from one site to another in various unobserved stages.
Each stage is exponentially distributed in time, but 
one observes only when $n$ stages have occurred. In this situation,
the non-Markovianity expressed by the non-exponential waiting time 
distribution is due to the lack of information
about all the intermediate stages \cite{Breuer2008}.
Indeed, one expects that the higher is the number of 
unobserved stages the stronger is the deviation from the Markovian
situation. Now, $\mathcal{N}_C(\Lambda)$ allows to formulate this
statement in a precise way since it quantifies the memory effects in the 
evolution of ${\bi{p}}(t)$,
which represent a signature of non-Markovianity. 
In figure~1, we plot $\mathcal{N}_C(\Lambda)$ for a semi-Markov process
that describes a system moving between two different sites with a 
waiting
time distribution as in (\ref{conv}).
\begin{figure}\label{fig:1}
\begin{center}
   \includegraphics[scale=0.40]{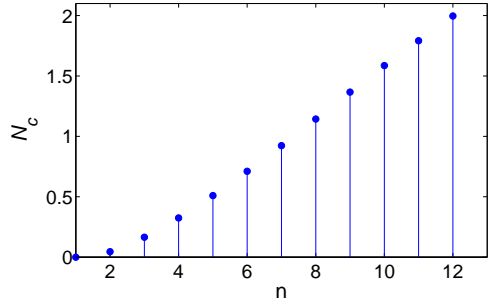}
\end{center}\caption{Size of memory effects in the evolution of ${\bi{p}}(t)$ for  a 
two-site semi-Markov process versus the number $n$ of the
exponential distributions in the special Erlang waiting time 
distribution (\ref{conv}) that fixes the process.
The memory effects are quantified by means of $\mathcal{N}_C(\Lambda)$ defined in 
(\ref{eqq}),
where $q(t)$ is obtained through equation (\ref{eq:qu}). For $n=2$, 
$\mathcal{N}_C(\Lambda) = 0.045$ (compare with figure~\ref{fig:2}).}
\end{figure}
For $f(t) = \lambda e^{- \lambda t}$
the Kolmogorov distance is monotonically decreasing \cite{Vacchini2011}, 
so that $\mathcal{N}_C(\Lambda) = 0$. Then,
$\mathcal{N}_C(\Lambda)$ increases with the number of exponential
distributions in (\ref{conv}) and, quite remarkably, 
it grows to a good approximation in a linear way
for $n > 3$. The memory introduced  by the lack of information about 
intermediate
stages growths linearly with the number 
of 
unobserved stages. In addition,
let us note that $\mathcal{N}_C(\Lambda)$ does not depend on the parameter 
$\lambda$ in (\ref{conv}).
By the properties of the Laplace transform or by dimensional analysis,
one can easily find
that $q(t)$ is a function of $\lambda t$. 
Maxima and minima of $q(t)$ do not depend on $\lambda$ and thus
neither does $\mathcal{N}_C(\Lambda)$, see the
right hand side of (\ref{eqq}). 
Only the functional expression of the waiting time distribution matters.

As a further example, we consider the convolution of two non-exponential
waiting time distributions. Coming back to the picture of a system 
jumping between
two sites with unobserved stages, we study how the memory in the 
evolution
of ${\bi{p}}(t)$ is influenced by the specific waiting time distribution
that characterizes the intermediate stages.
Let the waiting time distribution $f(t)$ be the convolution of two equal mixtures of exponential distributions 
with rates, respectively, $\lambda_1$ and $\lambda_2$: given $h(t)\equiv\mu \lambda_1 e^{- \lambda_1 t}+(1-\mu) \lambda_2 e^{-\lambda_2 t}$, we set
\begin{equation}\label{eq:mixx}
f (t)= (h  \ast h)(t). 
\end{equation} 
A mixture $\sum_{k} \mu_k f_k(t)$ describes a
system that moves between different sites following with probability 
$\mu_k$ the distribution $f_k(t)$.
Thus, $f(t)$ as in (\ref{eq:mixx}) means that 
both the unobserved stages are exponentially distributed in time with a 
rate that is given by $\lambda_1$, with probability $\mu$, or by 
$\lambda_2$, with probability $1-\mu$.
\begin{figure}\label{fig:2}
\begin{center}
   
\includegraphics[scale=0.44]{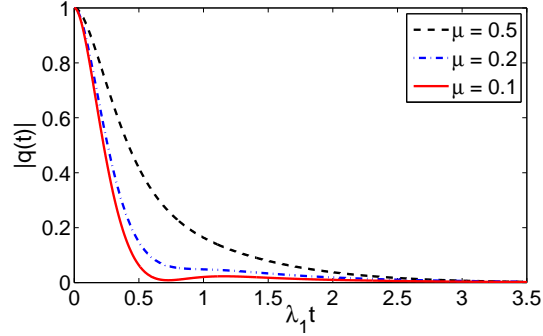}
\end{center}\caption{Plot of $|q(t)|$ versus time for a waiting time 
distribution given by the convolution of two equal mixtures of 
exponential distributions, see (\ref{eq:mixx}), for different
values of the mixing parameter $\mu$; in addition, $\lambda_2 / 
\lambda_1 = 5$.
$\mathcal{N}_C(\Lambda)$ in (\ref{eqq}) is equal to $0$ for
$\mu=0.5$ and $\mu=0.2$, while $\mathcal{N}_C(\Lambda) = 0.022$ for $\mu$=0.1.}
\end{figure}
In figure~\ref{fig:2}, we plot the evolution of $|q(t)|$, see 
(\ref{eq:qu}), for $f(t)$ as in (\ref{eq:mixx}).
Recall that $|q(t)|$ is the Kolmogorov distance for the two initial 
probability
vectors that maximize $\mathcal{N}_C(\Lambda)$, see (\ref{eqq}). Moreover, if the waiting time distribution is simply
the mixture of two exponential distributions, then the 
Kolmogorov distance
is monotonically decreasing \cite{Vacchini2011}, so that 
$\mathcal{N}_C(\Lambda)=0$. 
We can see in figure~\ref{fig:2} that this is still the case for
the convolution of two mixtures with balanced exponential 
distributions, i.e., $\mu=1/2$.
Even more, only a strong imbalance between the weights of the two 
exponential distributions provides
an increase of the Kolmogorov distance, witnessing the presence of 
memory effects.
In any case, this signature of non-Markovianity is smaller than
for the convolution of two equal exponential distributions, i.e., for 
$f(t)$ as in (\ref{conv})
with $n=2$, which is recovered from (\ref{eq:mixx}) for $\mu=0$ or 
$\mu=1$.
If the rate of the exponential 
distributions of the unobserved stages is not known with certainty, 
the memory due to the lack of  information about
the occurrence of the stages is reduced.
In particular, if the available exponential distributions are equally 
probable, there will be no memory effects at all.

\section{Kolmogorov distance and time-local equation}

In this section, we 
study the relation between the Kolmogorov distance and the differential 
equation satisfied by ${{\bi{p}}(t)}$,
which provides further signatures of non-Markovianity. From the 
knowledge 
of the dynamical maps $\Lambda(t, 0)$, one can directly obtain a 
time-local equation in the form
\begin{equation}\label{eq:tl}
\frac{\rmd}{\rmd t} {{\bi{p}}(t)} = L(t) {{\bi{p}}(t)},
\end{equation}
by means of the relation \cite{Hanggi1982, Chruscinski2011}
\begin{equation}
L(t) = \frac{\rmd  \Lambda(t, 0)}{\rmd t} \Lambda^{-1}(t, 0).
\end{equation}
In particular, consider the case such that $L(t)$
satisfies the so-called Kolmogorov conditions for any $t \geq0$, i.e., 
$(L(t))_{j k} \geq 0$ for $j \neq k$, $(L(t))_{j j} \leq 0$ and $\sum_k 
(L(t))_{j k} = 0$ for any $j$.
Equation (\ref{eq:tl}) is then equivalent to the system of equations
\begin{equation}\label{pauli}
\frac{\rmd}{\rmd t} p_j(t) = \sum_k (W_{j k}(t) p_k(t) -W_{k j}(t) 
p_j (t))
\end{equation}
with $W_{j k}(t) \geq 0$ linked to $(L(t))_{j k}$ through 
$(L(t))_{j k} = W_{j k}(t) - \sum_l W_{l k}(t) \delta_{j k}$.
One can show \cite{Norris1999} that $L(t)$ satisfies the Kolmogorov 
conditions for any $t\geq 0$
if and only if the corresponding family of dynamical maps 
$\left\{\Lambda(t, 0)\right\}_{t \geq0}$ satisfies
\begin{equation}\label{pdiv}
\Lambda(t, 0) = \Lambda(t, s) \Lambda(s, 0)
\end{equation}
for any $t\geq s \geq0$, with $\Lambda(t, s)$ itself being a stochastic 
matrix. 
The composition law (\ref{pdiv}), named P-divisibility in 
\cite{Vacchini2011}, represents the classical counterpart of the 
composition law
exploited in \cite{Rivas2010} to define the notion of quantum 
Markovianity \cite{Vacchini2011, Vacchini2012}.
The violation of P-divisibility provides
a further signature of the non-Markovianity of classical stochastic 
processes at the level of single-time probability distributions. 
Note that this signature of non-Markovianity can be directly read from 
the time-local equation (\ref{pauli}) since
P-divisibility is violated if and only if $W_{j k}(t) <0$  for some $j, 
k$ and $t\geq 0$. 
As emphasized \cite{Breuer2002, Maniscalco2004} for the quantum 
dynamics of statistical operators, also the evolution of the 
single-time probability distribution of a classical stochastic process 
can be described through
a time-local equation even if the process is non-Markovian. In 
particular, negative
coefficients in the equation will account for memory effects in the 
evolution 
of  ${\bi{p}}(t)$ \cite{Laine2012}.

For the two-site semi-Markov processes described in the previous 
section,
with $\Lambda(t, 0)$ as in (\ref{eq:kdp}), one has
\begin{equation}
L(t)= \left(\begin{array}{cc}
    -\gamma(t) & \gamma(t) \\
    \gamma(t)& -\gamma(t) 
  \end{array}\right),
\end{equation}
with
\begin{equation}\label{gamma}
 \gamma(t) = -\frac{\rmd q(t)/\rmd t}{2 q(t)}.
\end{equation}
Thus, the elements $p_j(t)$, $j=1,2$, satisfy the following system
of differential equations:
\begin{eqnarray}\label{eqc}
\frac{\rmd}{\rmd t} p_1 (t) &=& \gamma(t)\, (p_2(t)-p_1(t))\nonumber 
\\
\frac{\rmd}{\rmd t}p_2 (t) &=& \gamma(t)\, (p_1(t) -p_2(t)).
\end{eqnarray} 
The rate $\gamma(t)$ in  (\ref{eqc}) is positive if and only if the 
corresponding Kolmogorov distance is monotonically decreasing, see 
(\ref{kkdd}) and (\ref{gamma}).
For these processes the increase of the Kolmogorov distance and the 
negativity of the coefficients in the time local equation (\ref{pauli}),
or, equivalently, the violation of P-divisibility, provide signatures of 
non-Markovianity that are fully equivalent.
One could show that this is the case for any two-site semi-Markov 
process with a site-independent waiting time distribution.

Now, consider any non-Markov process whose single-time probability 
distribution satisfies
\begin{eqnarray}\label{eqcex}
\frac{\rmd}{\rmd t} p_1 (t) &=& \gamma_2(t) p_2(t)- \gamma_1(t) 
p_1(t)\nonumber \\
\frac{\rmd}{\rmd t}p_2 (t) &=& \gamma_1(t) p_1(t) - \gamma_2(t) 
p_2(t),
\end{eqnarray} 
with 
\begin{eqnarray}
\gamma_2(t) <0 \quad \mbox{for some}\,\, t\geq0; \quad \gamma_1(t) \geq 
0 \quad \forall\,\,t\geq0 \nonumber\\
\gamma_1(t) + \gamma_2(t) \geq0 \quad \forall\,\,t \geq0.\label{cond}
\end{eqnarray}
If we further ask $\int^t_0 \rmd \tau \gamma_2(\tau) \geq 0$ for any 
$t \geq0$ the resulting dynamics
is well-defined, i.e. the positivity of ${\bi{p}}(t)$ is conserved. By 
solving
(\ref{eqcex}), one easily finds
\begin{equation}
D_K \left({\bi{p}}^1(t), {\bi{p}}^2(t)\right) = e^{- \int^t_0 \rmd 
\tau (\gamma_1(\tau) + \gamma_2(\tau))} D_K \left({\bi{p}}^1(0), 
{\bi{p}}^2(0)\right),
\end{equation}
which is a monotonically decreasing function because of the last 
condition in (\ref{cond}).
Thus, there are processes such that the  Kolmogorov distance $D_K 
\left({\bi{p}}^1(t), {\bi{p}}^2(t)\right)$ does not increase in 
time, even if there is some negative coefficient
in the time-local equation (\ref{pauli}), see the first condition in 
(\ref{cond}), so that P-divisibility is violated:
these signatures of non-Markovianity are in general not equivalent.

\section{Conclusions}

We have investigated signatures of non-Markovianity in the evolution
of the single-time probability distribution ${\bi{p}}(t)$ of classical 
stochastic processes. We have first
focused on the increase of the Kolmogorov distance between probability 
vectors.
This allows to detect memory effects related to the initial condition in 
the subsequent evolution of ${\bi{p}}(t)$
and it naturally leads to the introduction of  a quantity that measures 
the relevance of the memory
effects, which can be seen as the classical counterpart of the measure 
for quantum non-Markovianity
introduced in \cite{Breuer2009}.
We have then discussed the relation between the increase of the 
Kolmogorov distance and the 
time-local equation satisfied by the probability vector. We have shown 
how for two-site
semi-Markov processes with a site-independent waiting time distribution 
the negativity of the rates 
in the time-local equation is equivalent to a non-monotonic evolution
of the Kolmogorov distance, further presenting examples
for which this is not the case.

\ack
We are grateful to Alberto Barchielli for many enlightening discussions. 
Financial support from MIUR under PRIN 2008 and the COST Action MP1006 
is acknowledged.

\end{document}